\def\cm{{\rm cm}$^{-1}$}

\documentclass[twocolumn,showpacs,preprintnumbers,amsmath,amssymb]{revtex4}

\usepackage{graphicx}
\usepackage{dcolumn}
\usepackage{bm}

\begin{document}

\preprint{Superlattices/PRB/Brief Report}

\title{Suppression of Superconductivity in YBCO/LCMO Superlattices}

\author{F. Chen$^{1,2}$, B. Gorshunov$^{1\dagger}$, G. Cristiani$^{2}$,
H.-U. Habermeier$^{2}$, and M. Dressel$^{1*}$}
\address{$^{1}$1.~Physikalisches Institut, Universit\"at Stuttgart, Pfaffenwaldring 57,
D-70550 Stuttgart, Germany\\
$^{2}$Max-Planck-Institut f\"ur Festk\"orperforschung, Heisenbergstr. 1, D-70569 Stuttgart, Germany}

\date{\today}

\begin{abstract}
The competition of superconductivity and magnetism in superlattices composed of alternating
YBa$_2$Cu$_3$O$_{7-d}$ and La$_{0.67}$Ca$_{0.33}$MnO$_{3}$ thin films is investigated using  low-energy
optical spectroscopy. The thickness of the superconducting YBCO layers is varied from 30~nm to 20~nm while
the thickness of the magnetic LCMO layers is kept constant at 20~nm. We clearly observe that the
superconducting condensate density in the superconducting state of superlattice is drastically reduced by
the magnetic subsystem which may be connected with proximity effects that distort the gap symmetry and
thus suppress superconductivity.
\end{abstract}


\maketitle

Since the early 1960s the interplay of superconductivity (SC) and ferromagnetism (FM) has continuously drawn
attention \cite{Fisher73}. At first glance SC and FM exclude each other because the magnetic exchange field
breaks the Cooper pairs. Nevertheless the possible coexistence of both phenomena due to a spatially modulated
order parameter was suggested by Fulde, Ferrell, Larkin and Ovchinnikov \cite{Fulde64}; the correspondent state
was never observed in conventional SC. A good way to study the effect is to bring the nano-sized layers of the two
materials into contact and to view the SC-FM interaction on the level of the proximity effects. The exploration of
multilayer systems composed of alternating conventional SC and metallic FM layers was subject of recent reviews
\cite{Chien99,Izyumov02}.

In the last years artificially grown superlattices (SLs) consisting of high-$T_c$ superconductors and
manganese oxides which exhibit the phenomenon of colossal magnetoresistance have attracted increasing
interests for various reasons. On one hand, SLs were developed as an important tool to explore the
interplay between the two antagonistic SC and FM ground states, and on the other hand the injection of
spin-polarized carriers can lead to new SC switching devices \cite{Goldman01}. Due to the structural
similarity of the two classes of perovskite compounds, it is possible to construct a unique combination of
SC cuprate and strongly FM manganese layers. It has been demonstrated by different groups
\cite{Jakob95,Przyslupski99,Habermeier01,Prieto01} that although SC and FM are preserved in each
subsystem, the superconducting transition temperature $T_c$ and the Curie temperature of the magnetic
ordering $T_{\rm mag}$ are considerably suppressed. Transport and magnetization measurements clearly
indicate that spin-polarized quasiparticle injection in
YBa$_2$Cu$_3$O$_{7-d}$/La$_{0.67}$Ca$_{0.33}$MnO$_3$ (YBCO/LCMO) SL can strongly decrease $T_c$ and the
critical current $J_c$ \cite{Vasko97,Yeh99}. Particularly interesting is the almost complete spin
polarization in LCMO and the low carrier density in YBCO which seems to be suppressed even further in the
case of SL, as was shown by optical spectroscopy \cite{Holden03}. The advantage of optical techniques is
that one can directly monitor the temperature dependence of the ``strength'' of the conducting and
superconducting condensates expressed by the spectral weight, i.e.\ the area under the dynamical
conductivity spectrum. Up to now, however, optical measurements failed to even detect a sign of (weakened)
superconductivity in SL. In this paper, we report on the first measurements of optical properties of
YBCO/LCMO superlattices at the radiation quantum energies low enough to allow for observation of SC
condensate response. We find strongly reduced SC superfluid density and an enhanced London penetration
depth in the superlattices and ascribe it to proximity effects.


\begin{figure}
\includegraphics[width=70mm]{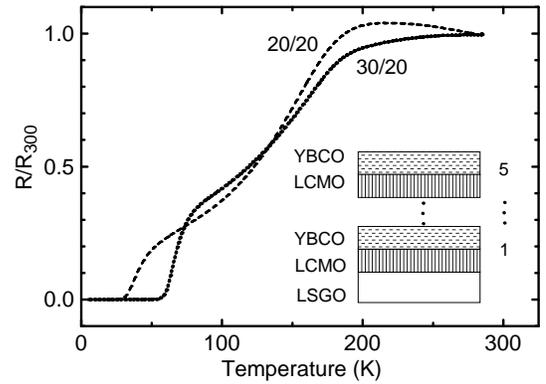}
\caption{\label{fig1}Temperature dependence of normalized dc-resistivities of two superlattices deposited
on a LaSrGaO$_4$ substrate, each consisting of five periods of 20~nm thick La$_{0.67}$Ca$_{0.33}$MnO$_3$
layer and 20 or 30~nm YBa$_2$Cu$_3$O$_{7-d}$ layer, as shown in the inset.}
\end{figure}
The samples are fabricated by pulsed laser deposition techniques on LaSrGaO$_4$ (LSGO) (001) substrates
which has low dielectric losses ($\tan\delta = 5\times 10^{-5}$) in the used frequency range and a
favorable in-plane lattice match between LSGO ($a=b=0.3840$~nm) and YBCO ($a=0.3836$~nm, $b=0.3883$~nm).
The growth conditions of the $c$-axis oriented YBCO and LCMO layers were optimized in the way to achieve
optimal superconductivity and ferromagnetism. Detailed information about the  fabrication of SLs can be
found in Ref.~\cite{Habermeier01}. Alternating LCMO and YBCO layers (five of each) are deposited as
sketched in the inset of Fig.~\ref{fig1}. For all SLs used, the thickness $d$ of the LCMO layers is kept
at 20~nm, while the thickness of the YBCO layers changes from 20~nm to 30~nm. We have studied two
superlattices, which will be referred to as 20/20 and 30/20 - the ratio of thicknesses of the YBCO to the
LCMO layers. Parameters of the samples used in our study are listed in Table~\ref{table1}. Only the
corresponding ($00h$) peaks are observed in X-ray diffraction $\theta/2$ measurements for YBCO, LCMO, and
the LSGO substrate. The SC transition temperature $T_c$ and the magnetic transition temperature $T_{\rm
mag}$ are determined by dc-resistance $R(T)$ and SQUID magnetization measurements, respectively
(Fig.~\ref{fig1}); the values are close to what we have obtained for SLs grown on SrTiO$_3$ substrate
\cite{Habermeier01,Holden03}. In addition, single LCMO film ($d=20$~nm) and single YBCO film ($d=30$~nm)
were also prepared on LSGO substrates so that we could study optical properties of thin YBCO and LSCO
films not incorporated in the SL. For a single thin YBCO film of 30~nm we find a reduced SC transition
temperature of approximately 70 to 80~K.

\begin{table}
\caption{\label{table1}Parameters of superlattices: thicknesses $d_{\rm YBCO}$ of YBa$_2$Cu$_3$O$_{7-d}$
and $d_{\rm LCMO}$ of La$_{0.67}$Ca$_{0.33}$MnO$_{3}$ films, number of film periods deposited,
superconducting transition temperature $T_c$, magnetic ordering temperature $T_{\rm mag}$ determined by
maximum of resistivity, London penetration depth $\lambda_L$, plasma frequency of superconducting
condensate $\omega^s_p/2\pi$, and scattering rate of normal carriers $\gamma$. Also listed are the
literature values for single crystal and thin film of YBa$_2$Cu$_3$O$_{7-d}$ for comparison.}
\begin{ruledtabular}
\begin{tabular}{cccccccc}
$d_{\rm YBCO}$ & $d_{\rm LCMO}$ & periods & $T_c$ & $T_{\rm mag}$
& $\lambda_L$& $\omega^s_p/2\pi$ &$\gamma$\\
 &  & & &  & (5~K) &(5~K)& (5~K) \\
(nm) & (nm) & &(K) & (K) & ($\mu$m) &(\cm)& (\cm) \\
\hline
30 & & 1 & 70 & & 0.9 & 1780 &12\\
 &20 & 1 &  & 245&  & & \\
30 &20 & 5 & 55 & 150& 1.9 & $840$&$>200$\\
20 &20 & 5 & 25 & 160 & 4.0 & $400$&$>200$\\
\hline
\multicolumn{3}{l}{YBCO crystal (Ref.~[\protect\onlinecite{Hardy93}])}  & 92& &0.14   &11370\\
\multicolumn{3}{l}{YBCO film  (Ref.~[\protect\onlinecite{Djordjevic98}])}    & 90& &0.20 &7960\\
\end{tabular}
\end{ruledtabular}
\end{table}

Optical measurements were done in transmission mode at the lower end of the far-infrared spectral range (6
- 30~\cm) at temperatures $5~{\rm K}< T < 300$~K, with the help of the millimeter-submillimeter
coherent-source spectrometer utilizing backward-wave oscillators as monochromatic frequency-tunable
radiation sources, as described elsewhere in detail \cite{Kozlov98}. In a Mach-Zehnder arrangement we
measure the transmission coefficient {\it Tr} and the phase shift $\varphi_t$ of the radiation passing
through the plane-parallel substrate with the SL on it. Spectra of both, conductivity  $\sigma(\omega)$
and dielectric permittivity $\epsilon(\omega)$, of single films or of composite SL are obtained directly
(no Kramers-Kronig relations used) from {\it Tr} and $\varphi_t$ on the basis of the general Fresnel
expressions for the layered system \cite{DresselGruner} (the dielectric parameters of the substrate are
measured beforehand). In these measurements the probing radiation passes through the superlattice which is
considered as a single layer and whose {\em effective} optical properties are thus averaged over those of
SC and FM layers composing it, in conditions when the radiation wavelength is much larger compared to the
layer thickness; we note that this {\em transmission} technique is thus more sensitive to the optical
properties of the films when compared to the optical reflectivity techniques.


\begin{figure}
\includegraphics[width=80mm]{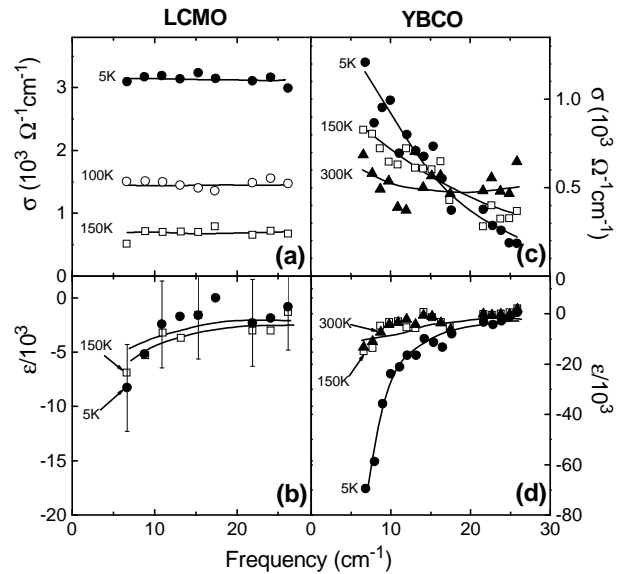}
\caption{\label{fig2}Frequency dependence of conductivity $\sigma(\omega)$ and dielectric permittivity
$\epsilon(\omega)$ of (a and b) single La$_{0.67}$Ca$_{0.33}$MnO$_3$ (thickness $d=20$~nm) and (c and d)
single YBa$_2$Cu$_3$O$_{7-d}$ ($d=30$~nm) films on a LaSrGaO$_4$ substrate at different temperatures as
indicated. The  lines are guides to the eye.}
\end{figure}
Fig.~\ref{fig2} shows the frequency dependence of the conductivity $\sigma(\omega)$ and dielectric
permittivity $\epsilon(\omega)$ of single LCMO and YBCO films recorded at different temperatures. For the
LCMO film, basically no frequency dependence is observed for $\sigma$ whose value increases at lower
temperatures, the spectra of $\epsilon$ show a slight increase to high frequencies and do not
significantly depend on the temperature - both observations indicate good metallic (Drude-like)
properties. In the case of YBCO, displayed in Fig.~\ref{fig2}c and d, both $\sigma$ and $\epsilon$ are
frequency independent at high temperatures (metallic behavior) but reveal a pronounced dispersion at 5~K.
A strong increase of $\sigma$ is seen towards the lowest frequencies, similar to that observed in all
high-Tc's and connected to uncondensed (normal) carriers \cite{Tanner92,Bonn96}; such dispersion leads to
a peak in the temperature dependence of the conductivity, in case of our film - around 50~K for the
frequency of 25~\cm. The dielectric permittivity $\epsilon$ shows in the SC state a divergent dispersion
$\epsilon \propto -(1/\omega)^2$ which represents the dielectric response of the zero-frequency
delta-function in the conductivity spectrum, responsible for the infinite DC conductivity
\cite{DresselGruner}. Both behaviors of $\sigma$ and $\epsilon$ of the YBCO film clearly display the
appearance of a SC state. We have analyzed the 5 K spectra of YBCO film modeling the response of the
condensed SC carriers as $\epsilon_s(\omega)=-(\omega^s_p/\omega)^2$ and a Drude expressions
\cite{DresselGruner} for $\sigma(\omega)$ and $\epsilon(\omega)$ for uncondensed carriers contribution;
the results are listed in Table~\ref{table1}. For the plasma frequency of the superconducting carriers at
$T=5$~K we obtain $\omega^s_p/2\pi =1780$~\cm\ from which we get the value of the London penetration depth
$\lambda_L=c/\omega^s_p=0.9~\mu$m ($c$ is the velocity of light). Both values are significantly different
from those obtained for single crystals and thicker films \cite{Hardy93,Djordjevic98}: the value of
$\lambda_L$ is strongly enhanced and $\omega^s_p$ decreased, probably due to the effect of the film
thickness, which will be discussed in a subsequent publication.

\begin{figure}
\includegraphics[width=70mm]{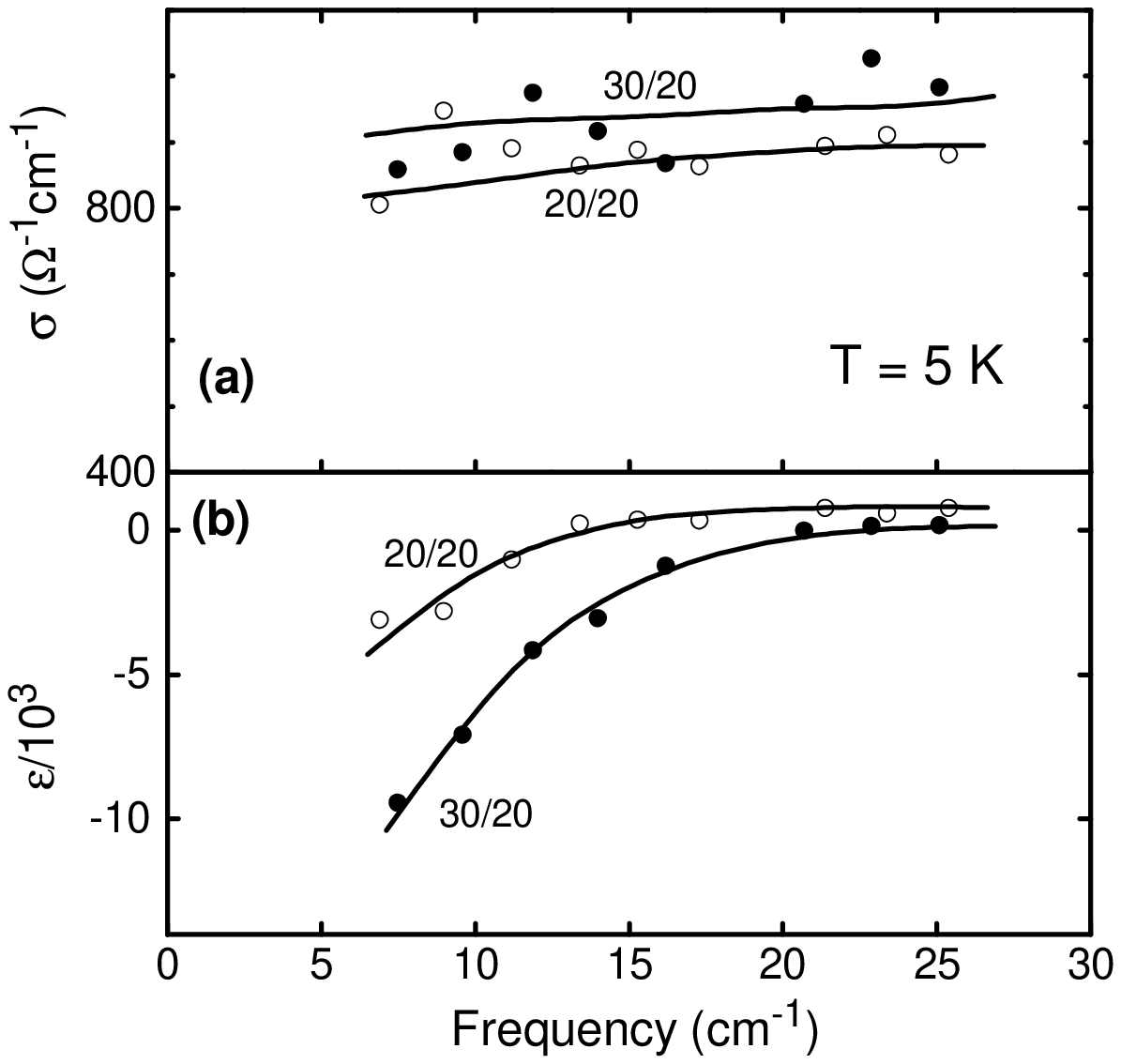}
\caption{\label{fig3}Frequency dependence of (a) conductivity $\sigma(\omega)$ and (b) dielectric
permittivity $\epsilon(\omega)$ of YBa$_2$Cu$_3$O$_{7-d}$/ La$_{0.67}$Ca$_{0.33}$MnO$_3$ super\-lattices
(five periods) with 20~nm thickness of the LCMO film and different thickneses of the YBCO film (20 and
30~nm, denoted as 20/20 and 30/20, respectively), shown for the temperature of $T = 5$~K. The solid lines
are guides to the eye. }
\end{figure}
\begin{figure}
\includegraphics[width=70mm]{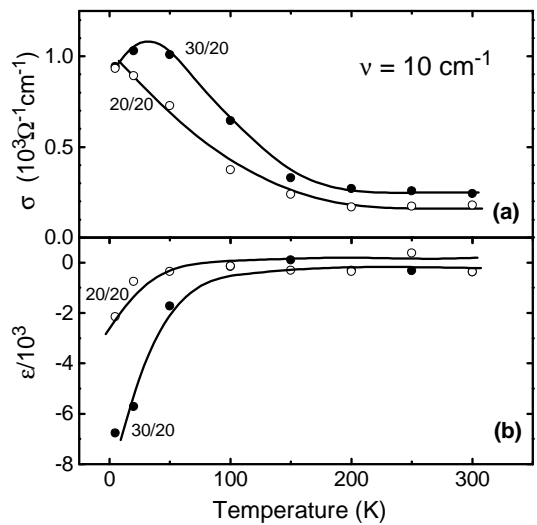}
\caption{\label{fig4}Temperature dependence of (a) conductivity $\sigma(T)$ and (b) dielectric
permittivity $\epsilon(T)$ of YBa$_2$Cu$_3$O$_{7-d}$/ La$_{0.67}$Ca$_{0.33}$MnO$_3$ super\-lattices (five
periods) with 20~nm thickness of the LCMO film and different thickness of the YBCO film (20 and 30~nm,
denoted as 20/20 and 30/20, respectively). The data are shown for one particular frequency $\nu = 10$~\cm.
The solid lines are guides to the eye.}
\end{figure}
We now turn to the dielectric responses of the superlattices for which the frequency and temperature
dependences are shown in Figs.~\ref{fig3} and \ref{fig4}. Similar to the single YBCO film, the dielectric
permittivity spectra of SLs also reveal a signature of a superconducting response (Fig.~\ref{fig3}) -- a
diverging behavior at the lowest temperature, $\epsilon(\omega) \propto -(1/\omega)^2$. At the same time
the absolute values of $\epsilon$, which represent a measure of the SC condensate density, are noticeably,
by 6 to 10 times, reduced compared to a single YBCO film. This means that while the superconductivity is
still maintained in SLs, its ``strength'' is strongly suppressed. By fitting the spectra of
$\sigma(\omega)$ and $\epsilon(\omega)$ for a SLs as described above we obtain the plasma frequencies of
the SC condensate and the London penetration depths, the values given in Table~\ref{table1}
\cite{remark2}. We find a strong decrease of the SC carriers density ($\omega^s_p$) and a correspondent
increase of $\lambda_L$ when going from the single YBCO film to the 30/20 and further to the 20/20
superlattice.

The conductivity spectra of SLs also show behaviors different from that in the single YBCO film. The
absolute values of the conductivities for the 20/20 and 30/20 SLs are reduced and no significant
dispersion is observed, both indicating an increase of the scattering rate for the normal (uncondensed)
carriers.

The observed suppression of SC in the superlattices cannot be caused by such effects as incomplete
oxygenation or surface disorder, as was shown in \cite{Habermeier01} and \cite{Sefrioui2002}. We also note
that we were not able to fit the measured spectra of transmission coefficient and phase shift by
considering the SL as consisting of 5 YBCO and 5 LCMO layers each having properties of single YBCO and
LCMO films. This means that these films, when incorporated in a SL, cannot be considered as independent
layers. We thus conclude that it is the interaction of FM and SC condensates that is responsible for the
strong reduction of the observed SC carrier density. Our findings are in accordance with estimations of
the spin diffusion length of approximetely 30~nm \cite{Soltan03}.

In the study of impurity doped YBCO, the normalized critical temperature $T_c/T_{c0}$ ($T_{c0}$
corresponds to the transition temperature of a high quality YBCO) is found to show the dependence on the
normalized zero temperature superfluid density $n_s/n_0$ \cite{Ulm95,Franz97} which is quite different
from that predicted by a standard Abrikosov-Gor'kov theory \cite{Abrikosov}. Using $T_{c0}$=92 K (YBCO
single crystal), we found that the ratio $T_c/T_{c0}$ for our superlattices shows behaviors similar to
that in Zn- and Ni-doped YBCO thin films [24,25]; those results were explained by a gapless
superconductivity. In fact for low-$T_c$ SC/FM superlattices Sun {\em et al.} \cite{Sun02} recently
proposed that gapless superconductivity appears in both FM and SC regions near the interface. In the
superlattices, the influence of the FM layers upon the SC layers is expressed via a strong plane
perturbation, magnetic exchange field and proximity effect. As first pointed out by Abrikosov and Gor'kov
\cite{Abrikosov}, gapless superconductivity is a common feature in SC in the presence of strong
perturbation. In the gapless region, the electronic properties are drastically modified and lead, for
example, to a linearly temperature dependent specific heat, reduced $T_c$ and enhanced London penetration
depth \cite{Maki69}, in accordance with our optical observations.

In summary, from our low-energy optical experiments we find a strong weakening of the superconducting response of
superconducting/ferromagnetic YBCO/LCMO superlattices when the thickness of the YBCO SC layer in the superlattice
decreases: considering the superlattice as an effective medium we find a strong decrease of the effective density
of the superconducting condensate and correspondent enhancement of the London penetration depth. These results
indicate a strong suppression of superconductivity in YBCO/LCMO superlattices which can be attributed to the
pair-breaking effect of the ferromagnetic layers.


\begin{thebibliography}{99}
\bibitem{remark2}Although there is a clear tendency of the values with increasing thickness
$d_{\rm YBCO}$, a spread in the sample quality, in particular the interfaces, leads to some scattering of the
data.
\end{thebibliography}
\end{document}